\documentclass[preprint,superscriptaddress,onecolumn,showpacs,showkeys]{revtex4-1}
\usepackage{textcomp}
\usepackage{eurosym}
\usepackage{amsfonts}
\usepackage{array}
\usepackage{amsthm}
\usepackage{bm}
\usepackage{palatino}
\usepackage[colorinlistoftodos]{todonotes}
\usepackage{mathpazo}
\usepackage{supertabular}
\usepackage{subfig}
\usepackage{amssymb}
\usepackage{eurosym}
\usepackage{amsmath}
\usepackage{epsfig}
\usepackage{graphics}
\usepackage{color}
\usepackage{graphicx}
\usepackage[font={footnotesize,it}]{caption}
\usepackage[colorlinks=true,
            linkcolor=red,
            urlcolor=blue,
            citecolor=green]{hyperref}
\usepackage{graphicx,color}

\def\LOGO{%
\begin{picture}(0,0)\unitlength=1cm
\put (6,-0.9) {\includegraphics[width=5em]{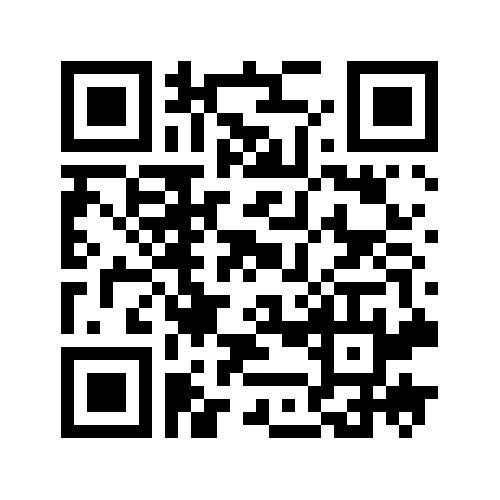}}
\end{picture}
}

\begin{document}

\begin{center}
  \sffamily\bfseries
  {\Large }\LOGO
\end{center}

\setcounter{MaxMatrixCols}{10}

\title{Absorption Cross-Section and Decay Rate of  Dilatonic Black Strings}
\author{Huriye G\"{u}rsel}
\email{huriye.gursel@emu.edu.tr}
\author{\.{I}zzet Sakall{\i}}
\email{izzet.sakalli@emu.edu.tr}
\email{izzet.sakalli@gmail.com}

\begin{abstract}
We studied in detail the propagation of a massive tachyonic scalar field in
the background of a five-dimensional ($5D$)
Einstein--Yang--Mills--Born--Infeld--dilaton black string: the massive
Klein--Gordon equation was solved, exactly. Next we obtained complete
analytical expressions for the greybody factor, absorption cross-section,
and decay-rate for the tachyonic scalar field in the geometry under
consideration. The behaviors of the obtained results are graphically
represented for different values of the theory's free parameters. We also
discuss why tachyons should be used instead of ordinary particles for the
analytical derivation of the greybody factor of the dilatonic $5D$ black
string.
\end{abstract}

\maketitle
\tableofcontents

\affiliation{Physics Department, Faculty of Arts and Sciences, Eastern
Mediterranean University, Famagusta, North Cyprus, via Mersin 10, Turkey}

\affiliation{Physics Department, Faculty of Arts and Sciences, Eastern
Mediterranean University, Famagusta, North Cyprus, via Mersin 10, Turkey}

\section{Introduction}

A wealth of information about quantum gravity can be obtained by studying
the unique and fascinating objects known as black holes (BHs). In BH
physics, greybody factors (GFs) modify black-body radiation, or predicted
Hawking radiation \cite{SHR1,SHR2}, within the limits of geometrical optics 
\cite{Optics}. In other words, GFs modify the Hawking radiation spectrum
observed at spatial infinity (SI), so that the radiation is not pure
Planckian \cite{Tez}.

GF, absorption cross-section (ACS), and decay-rate (DR) are quantities dependent upon
both the frequency of radiation and the geometry of spacetime. Currently,
although there are many studies of GF, ACS, and DR (see for example \cite%
{GF1,GF2,GF3,GF4,GF5,GF6} and the references therein), the number of
analytical studies of GFs that consider modified black-body radiation of
higher dimensional ($D>4$) spacetimes, like the BHs in string theory and
black strings \cite{BSs,BSs2,BSs3}, is rather limited (see for instance \cite%
{GF2,GF3,BS1,BS2,BS3,BS4,BS5}). This paucity of studies has arisen from the
mathematical difficulty of obtaining an analytical solution to the wave
equation of the stringent geometry being considered; in fact, analytical GF
computations apply to spacetimes in which the metric components are
independent of time. It is also worth noting that although BSs are defined
as a higher dimensional generalization of a BH, in which the event horizon
is topologically equivalent to $S^{2}\times S^{1}$ and spacetime is
asymptotically $M^{D-1}\times S^{1}$ four-dimensional ($4D$) BSs are also
derived.Lemos and Santos \cite{lem1,lem2,santos} showed that cylindrically
symmetric static solutions, with a negative cosmological constant, of the
Einstein--Maxwell equations admit charged $4D$ BSs. A rotating version of
the charged $4D$ BSs \cite{lem3,lem4} exhibits features similar to the
Kerr-Newman BH in spherical topology. The problem of analyzing GFs of scalar
fields from charged $4D$ BSs has recently been discussed by Ahmed and
Saifullah \cite{ahmed}. An interesting point about GFs has been reported by 
\cite{dof}: BH energy loss during Hawking radiation depends, crucially, on
the GF and the particles' degrees of freedom.

As mentioned above, further study of the GFs of BSs is required. To fill
this literature gap, in the current study, we considered dilatonic $5D$ BS 
\cite{Habib}, which is a solution to
Einstein--Yang--Mills--Born--Infeld--dilaton (EYMBID) theory. We
analytically studied its GF, ACS, and DR for massive scalar fields; however,
we considered tachyonic scalar particles instead of ordinary ones. The main
reason for this consideration is that using ordinary mass in the
Klein--Gordon equation (KGE) of the dilatonic $5D$ BS (as will be explained
in detail later) leads to the diverging of GFs. Roughly speaking, this is
due to the flux of the propagating waves of the ordinary massive scalar
fields. Namely, once the scalar fields to be considered belong to the
massive ordinary particles, the incoming SI flux becomes zero. The latter
remark implies that detectable radiation emitted from a dilatonic $5D$ BS
spacetime belongs to the massive tachyonic scalar fields. Therefore, the
current study focuses on the wave dynamics of tachyonic particles moving in
dilatonic $5D$ BS spacetime. However, using tachyonic modes in $5D$ geometry
should not be seen as nonphysical; instead they should be considered as the
imaginary mass fields rather than to faster-than-light particles \cite%
{mytachyon}. First, Feinberg \cite{Feinberg,Feinberg2} proposed that
tachyonic particles could be quanta of a quantum field with imaginary mass.
It was soon realized that excitations of \textit{such imaginary mass fields
do not in fact propagate faster than light} \cite{immass}. 

Following the idea of Kaluza--Klein \cite{KK}, any $4D$ physical trajectory
is the projection of higher dimensional worldlines. Effective $4D$
worldlines associated with massive particles are causality constrained to be
timelike. However, the corresponding higher dimensional worldlines need not
be exclusively timelike, which gives rise to a topological classification of
physical objects. In particular, elementary particles in a $5D$ geometry
should be viewed as tachyonic modes. The existence of tachyons in higher
dimensions has been thoroughly studied by Davidson and Owen \cite{PLB}.
Furthermore, the reader may refer to \cite{Horowitz} to understand tachyon
condensation in the evaporation process of a BS. To find the analytical GF,
ACS, and DR, we have shown how to obtain the complete analytical solution to
the massive KGE in the geometry of a dilatonic $5D$ BS.

Our work is organized as follows. Following this introduction, a brief
overview of the geometry of the dilatonic $5D$ BS is provided in Sec. 2.
Section 3 describes the KGE of the tachyonic fields in the dilatonic $5D$ BS
geometry; we present the exact solution of the radial equation in terms of
hypergeometric functions. In Sec. 4, we compute the GF and consequently the
ACS and DR of the dilatonic $5D$ BS, respectively. We then graphically
exhibit the results of the ACS and DR. Section 5 concludes with the final
remarks drawn from our study.

\section{Dilatonic $5D$ BS in EYMBID Theory}

$D\left( =d+1\right) $-dimensional action in the EYMBID theory is given by 
\cite{Habib}%
\begin{equation*}
I_{EYMBID}=-\frac{1}{16\pi G_{(D)}}\int_{\mathcal{M}}d^{d}x\sqrt{-g} \Bigg[ 
\mathcal{R}-\frac{4\left( \bigtriangledown \psi \right) ^{2}}{D-2}+
\end{equation*}

\begin{equation}
4\chi ^{2}e^{-b\psi }\left( 1-\sqrt{1+\frac{Fe^{2b}}{2\chi ^{2}}}\right) %
\Bigg] ,
\end{equation}

where $\psi $ is the dilaton field, $\chi $ denotes the Born-Infeld
parameter \cite{Born}, and $b=-\frac{4}{d-2}\alpha $ with the dilaton
parameter $\alpha =\frac{1}{\sqrt{d-1}}$. $G_{(D)}$ represents the $D$%
-dimensional Newtonian constant\textbf{\ }and its relation to its $4D$ form $%
\left( G_{(4)}\right) $ is given by  
\begin{equation}
G_{(D)}=G_{(4)}L^{D-4},  \label{G}
\end{equation}%
where $L$ is the upper limit of the compact coordinate $\left(
\int_{0}^{L}dz=L\right) $. Furthermore, $\mathcal{R}$ stands for the Ricci
scalar and $F=F_{\lambda \rho }^{(\overline{a})}F^{(\overline{a})\lambda
\rho }$ where the 2-form Yang-Mills field is given by 
\begin{equation}
F^{(a)}=dA^{(\overline{a})}+\frac{1}{2\sigma }C_{(\overline{b})(\overline{c}%
)}^{(\overline{a})}\left( A^{(\overline{b})}\wedge A^{(\overline{c})}\right)
,  \label{F}
\end{equation}%
with $C_{(\overline{b})(\overline{c})}^{(\overline{a})}$ and $\sigma $ being
structure and coupling constants, respectively. The Yang-Mills potential $%
A^{(a)}$is defined by following the Wu-Yang ansatz \cite{Wu}

\begin{equation}
A^{(\overline{a})}=\frac{Q}{r^{2}}\left( x_{i}dx_{j}-x_{j}dx_{i}\right) ,
\end{equation}

\begin{equation}
r^{2}=\underset{i=1}{\overset{d-1}{\sum }}x_{i}^{2},\text{ }2\leq j+1\leq
i\leq d-1,1\leq \overline{a}\leq (d-1)(d-2)/2,
\end{equation}

where $Q$ is the Yang-Mills charge. The solution for the dilaton is as
follows

\begin{equation}
\psi =-\frac{\left( d-2\right) }{2}\frac{\alpha \ln r}{\alpha ^{2}+1}.
\label{dil}
\end{equation}%
\ \ On the other hand, the line-element of the dilatonic $5D$ BS is given by 
\cite{Habib}

\begin{equation}
ds_{0}^{2}=-\frac{f(r)}{\beta }d\widetilde{t}^{2}+\frac{\beta dr^{2}}{rf(r)}%
+rd\widetilde{z}^{2}+\beta \left( d\theta ^{2}+\sin ^{2}\theta d\phi
^{2}\right) ,  \label{metric0}
\end{equation}%
{} where $f(r)=r-r_{+}$ and $\beta =\frac{4Q^{2}}{3}$. $r_{+}$ represents
the outer event horizon having the following $(d+1)-$dimensional form:

\begin{equation}
\frac{32}{L^{d-4}}\left( \frac{Q^{2}d}{d-1}\right) ^{\frac{d-2}{2}}=r_{+}^{%
\frac{d(d-2)+2}{d}}.  \label{hori}
\end{equation}

Because, in our case, $D=5$ $\left( \text{i.e., }d=4\right) $, the horizon
becomes

\begin{equation}
r_{+}=4\beta ^{\frac{2}{5}}=4.488Q^{\frac{4}{5}}.  \label{hor}
\end{equation}

After rescaling the metric (\ref{metric0})

\begin{equation}
ds^{2}=\frac{ds_{0}^{2}}{\beta }=-\frac{f(r)}{\beta ^{2}}d\widetilde{t}^{2}+%
\frac{dr^{2}}{rf(r)}+\frac{r}{\beta }d\widetilde{z}^{2}+d\theta ^{2}+\sin
^{2}\theta d\phi ^{2},
\end{equation}

and in sequel assigning $\widetilde{t}$ and $\widetilde{z}$ coordinates to
the new coordinates:

\begin{equation}
\widetilde{t}\rightarrow \beta t,\text{ \ \ \ \ \ }\widetilde{z}\rightarrow
\beta z,
\end{equation}

we get the metric that will be used in our computations:

\begin{equation}
ds^{2}=-f(r)dt^{2}+\frac{dr^{2}}{rf(r)}+\beta rdz^{2}+d\theta ^{2}+\sin
^{2}\theta d\phi ^{2}.  \label{metric}
\end{equation}

It is worth noting that the surface gravity \cite{SurGrav} of the dilatonic $%
5D$ BS can be evaluated by

\begin{equation}
\kappa ^{2}=-\frac{1}{2}\left. \triangledown ^{\rho }\Upsilon ^{\sigma
}\triangledown _{\rho }\Upsilon _{\sigma }\right\vert _{r=r_{+}},
\label{SURFACE}
\end{equation}%
in which $\Upsilon ^{\mu }$ represents the timelike Killing vector: 
\begin{equation}
\Upsilon ^{\mu }=[1,0,0,0,0].  \label{killing}
\end{equation}%
Then, Eq.(\ref{SURFACE}) results in

\begin{equation}
\kappa=\left. \frac{\sqrt{r}f^{\prime}}{2}\right\vert _{r=r_{+}}=\frac {%
\sqrt{r_{+}}}{2},  \label{SG}
\end{equation}
where the prime denotes the derivative with respect to $r$. Furthermore, the
associated Hawking temperature is expressed by

\begin{equation}
T_{H}=\frac{\kappa }{2\pi }=\frac{\sqrt{r_{+}}}{4\pi }.  \label{HR}
\end{equation}

It is important to remark that the Hawking temperature of the dilatonic BS
given in Eq. (40) of Ref. \cite{Habib} is incorrect. The authors of Ref. 
\cite{Habib} computed the Hawking temperature of the dilatonic $5D$ BS
considering the metric to be symmetric, which is not the case since $%
g_{tt}\neq \frac{1}{g_{rr}}$. Meanwhile, it is obvious that dilatonic $5D$
BS (\ref{metric}) has a non-asymptotically flat structure. Therefore, it
possesses a quasilocal mass \cite{BY1,BY2,BY3}, which can be computed as
follows

\begin{equation}
M_{QL}=\frac{1}{6}\beta r_{h}^{\frac{3}{2}}L=\frac{4}{3}\beta ^{\frac{8}{5}%
}L\cong 2.113Q^{\frac{16}{5}}L.
\end{equation}

Thus, the first law of thermodynamics is satisfied:

\begin{equation}
dM_{QL}=T_{H}dS_{BH},
\end{equation}

where $S_{BH}$ denotes the Bekenstein-Hawking entropy \cite{SurGrav}, which
takes the following form for the dilatonic $5D$ BS

\begin{equation}
S_{BH}=\frac{A_{H}}{4}=\frac{1}{4}\beta r_{h}\int_{0}^{\pi }\sin \theta
d\theta \int_{0}^{2\pi }d\phi \int_{0}^{L}dz=\pi \beta Lr_{h}.
\end{equation}

\section{Wave Equation of a Massive Scalar Tachyonic Field in Dilatonic $5D$
BS}

As the scalar waves being studied belong to the massive scalar tachyons, the
corresponding KGE is given by

\begin{equation}
\left[ \square -\left( i\mu \right) ^{2}\right] \Psi (t,\mathbf{r})=\left[
\square +\mu ^{2}\right] \Psi (t,\mathbf{r})=0.  \label{KG}
\end{equation}%
We chose ansatz as follows:

\begin{equation}
\Psi (t,\mathbf{r})=R(r)Y_{l}^{m}(\theta ,\phi )e^{ikz}e^{-i\omega t},
\label{ANSATZ}
\end{equation}%
where $Y_{l}^{m}(\theta ,\phi )$ is the usual spherical harmonics and $k$ is
a constant. After making straightforward calculations, we obtained the
radial equation as follows:

\begin{equation}
rf\frac{\overset{\cdot \cdot }{R}}{R}+\frac{\overset{\cdot }{R}}{R}\left( f+r%
\overset{\cdot }{f}\right) +\frac{\omega ^{2}}{f}-\frac{k^{2}}{\beta r}+\mu
^{2}-\lambda =0,  \label{radial}
\end{equation}%
where $\lambda =l(l+1)$ and a dot mark denotes a derivative with respect to $%
r$. Multiplying each term by $r\beta f(r)R(r)$ and using the ansatz $y=\frac{%
r_{+}-r}{r_{+}}$, which in turn implies $r=r_{+}-yr_{+}$, one gets

\begin{equation}
y(1-y)R^{\prime \prime }+(1-2y)R^{\prime }+\left[ \frac{\omega ^{2}}{yr_{+}}+%
\frac{k^{2}}{\beta (1-y)r_{+}}-\mu ^{2}+\lambda \right] R=0,  \label{radial2}
\end{equation}

where prime denotes derivative with respect to $y$. Setting

\begin{equation}
\left[ \frac{\omega ^{2}}{yr_{+}}+\frac{k^{2}}{\beta (1-y)r_{+}}-\mu
^{2}+\lambda \right] =\frac{A^{2}}{y}-\frac{B^{2}}{1-y}+C,
\end{equation}

one can obtain

\begin{eqnarray}
A &=&-\frac{\omega }{2\kappa },  \notag \\
B &=&\frac{ik}{2\kappa \sqrt{\beta }}, \\
C &=&\lambda -\mu ^{2}.  \notag
\end{eqnarray}

Equation (\ref{radial2}) can be solved by comparing it with the standard
hypergeometric differential equation \cite{Steg} which admits the following
solution:

\begin{eqnarray}
R &=&\xi _{1}\left( -y\right) ^{iA}\left( 1-y\right) ^{-B}F\left(
a,b;c;y\right) +  \notag \\
&&\xi _{2}\left( -y\right) ^{-iA}\left( 1-y\right) ^{-B}F\left( \alpha
,\varsigma ;\eta ;y\right) ,  \label{iz1}
\end{eqnarray}%
where

\begin{equation}
a=\frac{1}{2}\left( 1+\sqrt{1+4C}\right) +iA-B,
\end{equation}

\begin{equation}
b=\frac{1}{2}\left( 1-\sqrt{1+4C}\right) +iA-B,
\end{equation}

\begin{equation}
c=1+2iA.
\end{equation}%
and

\begin{equation}
\alpha =a-c+1,  \label{alpha}
\end{equation}

\begin{equation}
\varsigma =b-c+1,  \label{const2}
\end{equation}

\begin{equation}
\eta =2-c.
\end{equation}%
According to our calculations, to have a non-zero SI incoming flux [see Eq. (%
\ref{FFSI})] or non-divergent GF [see Eq.(\ref{GF})] , $\sqrt{1+4C} $ must
be imaginary. To this end, we must impose the following condition:

\begin{equation}
4\mu ^{2}>4\lambda +1,
\end{equation}

such that

\begin{equation}
\sqrt{1+4C}=i\tau,  \label{tau}
\end{equation}
where

\begin{equation}
\tau=\sqrt{4\mu^{2}-4\lambda-1},\tau\in%
\mathbb{R}
.  \label{TAU}
\end{equation}
To obtain a physically acceptable solution, we must terminate the outgoing
solution at the horizon, which can be simply done by imposing $\xi_{1}=0$.
Thus, the physical radial solution reduces to

\begin{equation}
R=\xi _{2}\left( -y\right) ^{-iA}\left( 1-y\right) ^{-B}F\left( \alpha
,\varsigma ;\gamma ;y\right) .  \label{NH}
\end{equation}%
It should be noted that checking the forms of Eq. (\ref{NH}) both at the
horizon and at SI is essential. Section 3 shows that both are needed for the
evaluation of the GF. For the near horizon (NH) where $y\rightarrow 0$, one
can state:

\begin{equation}
R_{NH}=\xi _{2}\left( -y\right) ^{-iA},  \label{rr}
\end{equation}

which implies that the purely ingoing plane wave reads

\begin{equation}
\psi _{NH}=\xi _{2}e^{-i\omega (\widehat{r}_{\ast }+t)}e^{ikz},  \label{WNH}
\end{equation}%
where

\begin{eqnarray}
r^{_{\ast }} &=&\int \frac{dr}{\sqrt{r}f}\text{ \ }\rightarrow \text{ \ \ }%
\widehat{r}_{\ast }=\lim_{r\rightarrow r_{+}}r^{_{\ast }}\simeq \frac{\ln
\left( -y\right) }{\sqrt{r_{+}}},  \notag \\
\widehat{r}_{\ast } &\simeq &\frac{1}{2\kappa }\ln \left( -y\right) \text{ \ 
}\Longrightarrow \text{\ \ }y=-e^{2\kappa \widehat{r}_{\ast }}.  \label{tor}
\end{eqnarray}

On the other hand, for $y\rightarrow \infty $, the inverse transformation of
the hypergeometric function is given by \cite{Lay}

\begin{align}
F\left( \alpha ,\varsigma ;\eta ;y\right) & =\left( -y\right) ^{-\alpha }%
\frac{\Gamma (\eta )\Gamma (\varsigma -\alpha )}{\Gamma (\varsigma )\Gamma
(\eta -\alpha )}\times   \notag \\
& F\left( \alpha ,\alpha +1-\eta ;\alpha +1-\varsigma ;1/y\right) +  \notag
\\
& \left( -y\right) ^{-\varsigma }\frac{\Gamma (\eta )\Gamma (\alpha
-\varsigma )}{\Gamma (\alpha )\Gamma (\eta -\varsigma )}\times   \notag \\
& F\left( \varsigma ,\varsigma +1-\eta ;\varsigma +1-\alpha ;1/y\right) ,
\label{izn1}
\end{align}%
which yields the following asymptotic solution

\begin{eqnarray}
R_{SI} &\backsimeq &\xi _{2}\left( -y\right) ^{-iA-B-\alpha }\frac{\Gamma
(\eta )\Gamma (\varsigma -\alpha )}{\Gamma (\varsigma )\Gamma (\eta -\alpha )%
}+  \notag \\
&&\xi _{2}\left( -y\right) ^{-iA-B-\varsigma }\frac{\Gamma (\eta )\Gamma
(\alpha -\varsigma )}{\Gamma (\alpha )\Gamma (\eta -\varsigma )}.  \label{SI}
\end{eqnarray}%
To express Eq.(\ref{SI})\ in a more compact form, let us perform the
following simplifications. Considering

\begin{equation}
-iA-B-\alpha =-\frac{1}{2}\left( 1+i\tau \right) ,  \label{s1}
\end{equation}%
together with

\begin{equation}
-iA-B-\varsigma =-\frac{1}{2}\left( 1-i\tau \right) ,  \label{s2}
\end{equation}%
\ \ \ \ and letting $x=-y$, the radial equation for $r\rightarrow \infty $
takes the form

\begin{equation}
R_{SI}=\frac{1}{\sqrt{x}}\left[ \xi _{2}x^{-\frac{i\tau }{2}}\frac{\Gamma
(\eta )\Gamma (\varsigma -\alpha )}{\Gamma (\varsigma )\Gamma (\eta -\alpha )%
}+\xi _{2}x^{\frac{i\tau }{2}}\frac{\Gamma (\eta )\Gamma (\alpha -\varsigma )%
}{\Gamma (\alpha )\Gamma (\eta -\varsigma )}\right] .  \label{s3}
\end{equation}%
One can express $x$ in terms of the tortoise coordinate at SI as

\begin{equation}
r^{_{\ast }}=\int \frac{dr}{\sqrt{r}f}\text{ \ }\rightarrow \text{ \ \ }%
\widehat{r}_{\ast }=\lim_{r\rightarrow \infty }r^{_{\ast }}\simeq -\frac{2}{%
\sqrt{r}},
\end{equation}%
such that

\begin{equation}
x=r-r_{+}\text{ \ \ \ }\left. x\right\vert _{r\rightarrow \infty }\simeq
r=4e^{-2\widehat{r}_{\ast }},
\end{equation}%
where $\widehat{r}_{\ast }=\ln r^{_{\ast }}$. Therefore, we have

\begin{equation}
R_{SI}=\frac{1}{\sqrt{r}}\left[ \Lambda _{1}e^{i\widehat{r}_{\ast }\tau
}+\Lambda _{2}e^{-i\widehat{r}_{\ast }\tau }\right] ,  \label{s4}
\end{equation}%
where

\begin{equation}
\Lambda _{1}=2^{-i\tau }\xi _{2}\frac{\Gamma (\eta )\Gamma (\varsigma
-\alpha )}{\Gamma (\varsigma )\Gamma (\eta -\alpha )},
\end{equation}

\begin{equation}
\Lambda _{2}=2^{-i\tau }\xi _{2}\frac{\Gamma (\eta )\Gamma (\alpha
-\varsigma )}{\Gamma (\alpha )\Gamma (\eta -\varsigma )}.
\end{equation}%
Thus, the asymptotic wave solution becomes

\begin{equation}
\psi _{SI}=\frac{e^{ikz}}{\sqrt{r}}\left[ \Lambda _{1}e^{i(\widehat{r}_{\ast
}\tau -\omega t)}+\Lambda _{2}e^{-i(\widehat{r}_{\ast }\tau +\omega t)}%
\right] .  \label{WSI}
\end{equation}

\section{Radiation of Dilatonic $5D$ BS}

\subsection{The Flux Computation}

In this section, we compute the ingoing flux at the horizon $\left(
r\rightarrow r_{+}\right) $ and the asymptotic flux for the SI region $%
\left( r\rightarrow \infty \right) $. The evaluation of these flux values
will enable us to calculate the GF and subsequently, the ACS and DR.

The NH-flux can be calculated via \cite{Fernando,Flux}%
\begin{equation}
\digamma _{NH}=\frac{A_{BH}}{2i}\left( \overline{\psi }_{NH}\partial r_{\ast
}\psi _{NH}-\psi _{NH}\partial r_{\ast }\overline{\psi }_{NH}\right) ,
\label{FNH}
\end{equation}%
which, after a few manipulations, can be written as

\begin{equation}
\digamma _{NH}=-4\pi \beta \left\vert \xi _{2}\right\vert ^{2}r_{+}.
\label{NH flux}
\end{equation}%
The incoming flux at SI is computed via

\begin{equation}
\digamma _{SI}=\frac{A_{BH}}{2i}\left( \overline{\psi }_{SI}\partial r_{\ast
}\psi _{SI}-\psi _{SI}\partial r_{\ast }\overline{\psi }_{SI}\right) .
\label{FFSI}
\end{equation}%
Having performed the steps to evaluate the derivatives with respect to the
tortoise coordinate, the incoming flux at SI takes the form

\begin{equation}
\digamma _{SI}=-4\pi \beta \left\vert \Lambda _{2}\right\vert ^{2}\tau .
\label{SI flux}
\end{equation}

It is important to remark that if we were dealing with the standard
particles rather than tachyons, $\tau $ (\ref{TAU}) would be imaginary i.e., 
$\tau \rightarrow i\tau $ and therefore SI incoming wave would lead this flux evaluation (\ref{FFSI}%
) to be zero. This would indicate the existence of a divergent GF. 

\subsection{ACS of Dilatonic $5D$ BS}

The GF of the dilatonic $5D$ BS is obtained by the following expression \cite%
{GF2,GF4}

\begin{equation}
\gamma^{l,k}=\frac{\digamma_{NH}}{\digamma_{SI}}=\frac{-4\pi\beta\left\vert
\xi_{2}\right\vert ^{2}r_{+}}{-4\pi\beta\left\vert \Lambda_{2}\right\vert
^{2}\tau},  \label{GF}
\end{equation}
which is nothing but

\begin{equation}
\gamma ^{l,k}=\frac{\left\vert \xi _{2}\right\vert ^{2}r_{+}}{\left\vert
\Lambda _{2}\right\vert ^{2}\tau }.  \label{GF2}
\end{equation}%
After a few manipulations, with the following (see \cite{Steg})

\begin{equation}
\left\vert \Gamma (iy)\right\vert ^{2}=\frac{\pi }{y\sinh \left( \pi
y\right) },  \label{gm1}
\end{equation}

\begin{equation}
\left\vert \Gamma (1+iy)\right\vert ^{2}=\frac{\pi y}{\sinh \left( \pi
y\right) },  \label{gm2}
\end{equation}

\begin{equation}
\left\vert \Gamma (\frac{1}{2}+iy)\right\vert ^{2}=\frac{\pi }{\cosh \left(
\pi y\right) },  \label{gm3}
\end{equation}%
Eq. (\ref{GF2}) can be presented as

\begin{equation}
\gamma^{l,k}=\frac{\kappa r_{+}}{\omega}\left( e^{\frac{2\pi\omega}{\kappa}%
}-1\right) \Xi,  \label{GF3}
\end{equation}
where

\begin{equation}
\Xi =\frac{e^{2\pi \tau }-1}{\left[ e^{\pi \left( \tau +\frac{\omega }{%
\kappa }-\frac{k}{\kappa \sqrt{\beta }}\right) }+1\right] \left[ e^{\pi
\left( \tau +\frac{\omega }{\kappa }+\frac{k}{\kappa \sqrt{\beta }}\right)
}+1\right] }.  \label{coefffffff}
\end{equation}%
To evaluate the ACS of the dilatonic $5D$ BS concerned, we follow the study
of \cite{Kanti}. Thus, one can get the ACS expression in $5D$ as follows:

\begin{equation}
\sigma ^{l,k}=\frac{4\pi \left( l+1\right) ^{2}}{\omega ^{3}}\gamma ^{l,k},
\label{C1}
\end{equation}%
which, in our case, becomes\textit{\ } 
\begin{equation}
\sigma ^{l,k}=\frac{4\pi \left( l+1\right) ^{2}\kappa r_{+}}{\omega ^{4}}%
\left( e^{\frac{2\pi \omega }{\kappa }}-1\right) \Xi .  \label{C2}
\end{equation}

Furthermore, one can also get the total ACS as follows \cite{tcs}

\begin{equation}
\sigma _{abs}^{Total}=\sum\limits_{l=0}^{\infty }\sigma ^{l,k}.  \label{gm4}
\end{equation}

In Fig. 1., the relationship between absorption ACS and frequency is
examined; the figure is drawn based on Eq. (\ref{C2}). In the high frequency
regime, all ACSs tend to vanish by following the same curve. Unlike the high
frequency regime, ACSs diverge in the low frequency regime as $\omega
\rightarrow 0$. As a final remark, negative $\sigma ^{l,k}$ behavior has not
been observed in our graphical analyzes, which means that superradiance does
not occur \cite{chandra}, as expected (as the dilatonic $5D$ BS (\ref{metric}%
) does not rotate).

\begin{figure}[h]
\centering
\includegraphics[scale=.45]{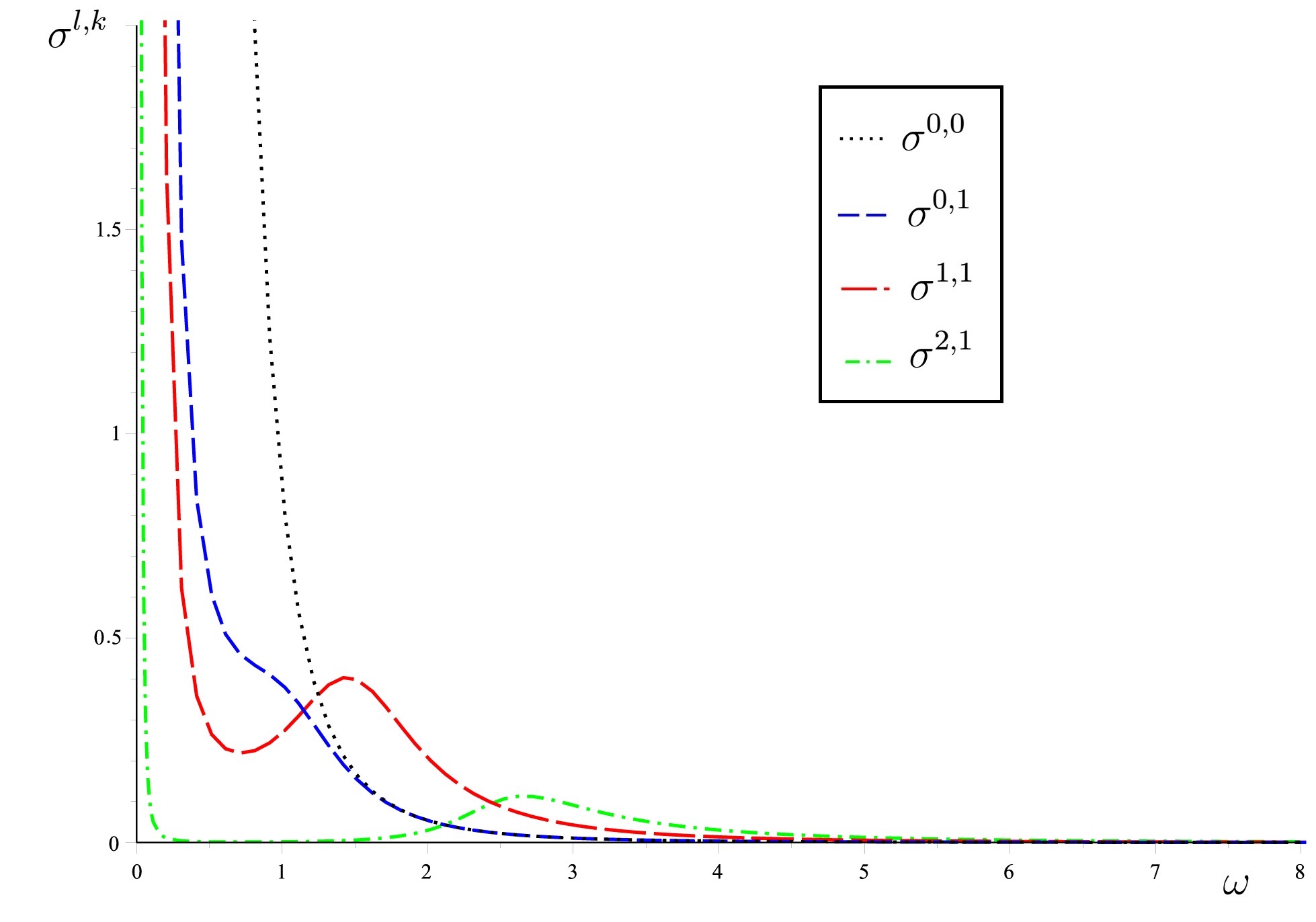}
\caption{Plots of the ACS ($\protect\sigma^{l,k}$) versus frequency $\protect%
\omega $. The plots are governed by Eq. (\protect\ref{C2}). The
configuration of the dilatonic $5D$ BS is as follows $\protect\mu=3$ and $%
Q=0.2$.}
\end{figure}

\subsection{DR of Dilatonic $5D$ BS}

The final step follows from the ACS evaluation. The DR of the dilatonic $5D$
BS can be computed via \cite{GF4}

\begin{equation}
\Gamma _{DR}^{l,k}=\frac{\sigma ^{l,k}}{e^{\frac{2\pi \omega }{\kappa }}-1}=%
\frac{4\pi \left( l+1\right) ^{2}\kappa r_{+}}{\omega ^{4}}\Xi .  \label{d}
\end{equation}

\begin{figure}[h]
\centering
\includegraphics[scale=.45]{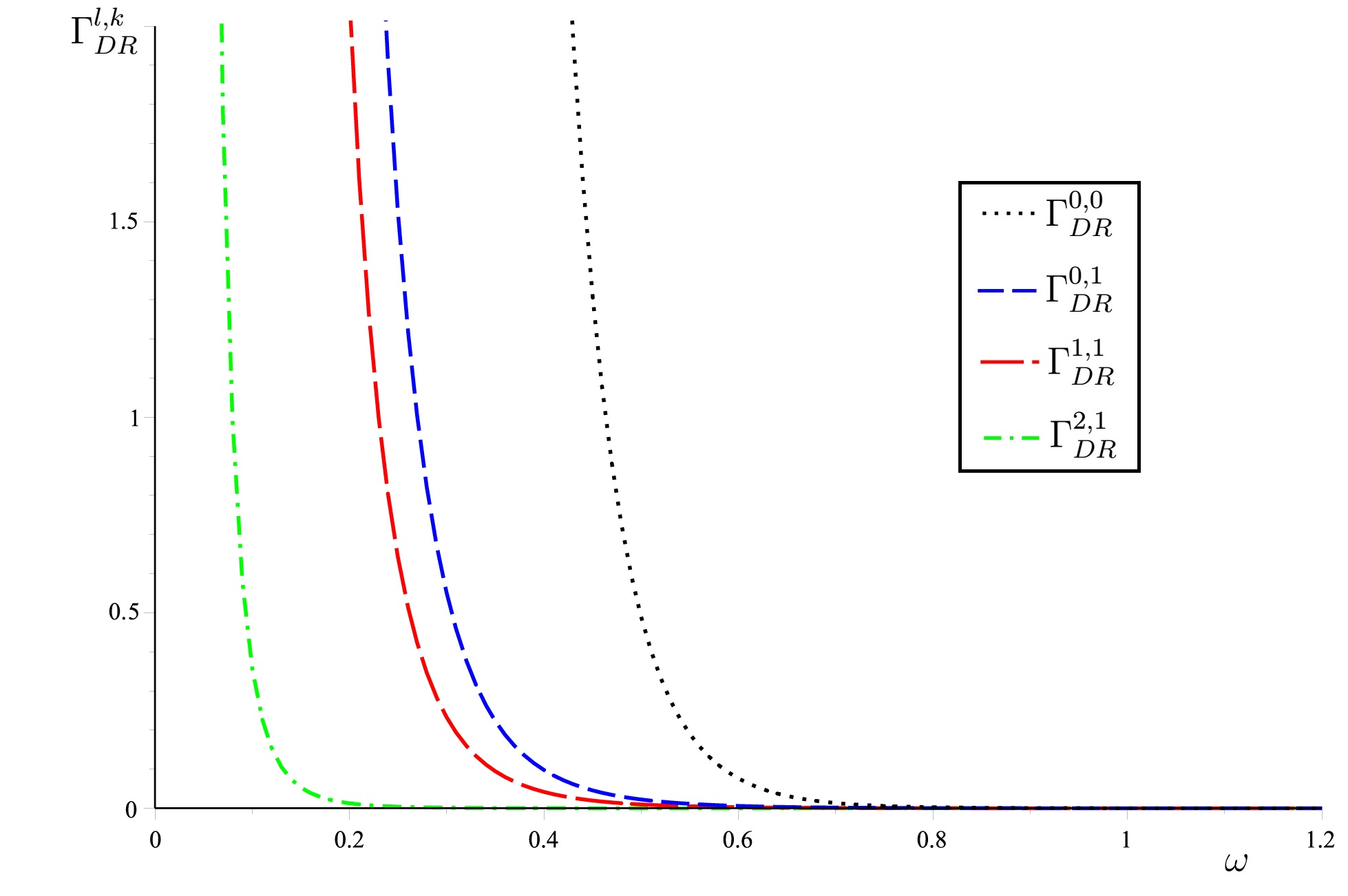}
\caption{Plots of the DR $(\Gamma_{DR}^{l,k})$ versus frequency $\protect%
\omega $. The plots are governed by Eq. (\protect\ref{d}). The configuration
of the dilatonic $5D$ BS is as follows $\protect\mu=3$ and $Q=0.2$.}
\end{figure}

Fig. 2. shows how the DR behaves with respect to the frequency. By taking
Eq. (\ref{d}) as the reference, the plots for increasing $l$ are
illustrated. In the high frequency regime, all DRs fade in the same way. In
the low frequency regime, DRs tend to diverge. However, it can be observed
that when $l$ has larger values, the corresponding DR diverges when $\omega $
is much closer to zero.

\section{Conclusion}

This article evaluated the GF, ACS, and DR for the dilatonic $5D$ BS geometry
arising from the EYMBID theory. As a result of the analytical method we
followed, it was shown that the radiation of the dilatonic $5D$ BS spacetime
can only be caused by tachyons. The crucial point here is that if standard
scalar particles had been used rather than tachyonic ones, zero incoming
flux at SI would have been obtained, which would lead to the diverging of
the GF. Therefore, in a way, we were forced to use tachyons to solve this
problem, and this carries great importance as it implies that the fifth
dimension could be directly linked to tachyons. In short, according to our
analytical method, we obtained results (compatible with boundary conditions)
when the radiation of the dilatonic $5D$ BS was provided by the tachyons.

In future study, we want to extend our analysis to the Dirac equation for
the geometry of the dilatonic $5D$ BS. Hence, we are planning to undertake
similar analysis for fermions and compare the results with scalar ones .

\section*{Acknowledgment}

We wish to thank Prof. Dr. Mustafa Halilsoy for drawing our attention to
this problem and for his helpful comments and suggestions. This work is supported by Eastern Mediterranean University through the project: BAPC-04-18-01.


\begin{thebibliography}{99}
\bibitem{SHR1} S. W. Hawking, Nature, \textbf{248,} 30 (1974).

\bibitem{SHR2} S. W. Hawking, Commun. Math. Phys., \textbf{43,} 199 (1975);
erratum: ibid, \textbf{46,} 206 (1976).

\bibitem{Optics} S. Creek, O. Efthimiou, P. Kanti, and K. Tamvakis, Phys.
Rev. D \textbf{75,} 084043 (2007).

\bibitem{Tez} P. Boonserm, \textit{Rigorous bounds on Transmission,
Reflection, and Bogoliubov coefficients}, arXiv:0907.0045 (2009).

\bibitem{GF1} C. Ding, S. Chen, and J. Jing, Phys. Rev. D \textbf{82},
024031 (2010).

\bibitem{GF2} T. Harmark, J. Natario, and R. Schiappa, Adv. Theor. Math.
Phys. \textbf{14,} 727 (2010).

\bibitem{GF3} P. Kanti, T. Pappas, and N. Pappas, Phys. Rev. D \textbf{90},
124077 (2014).

\bibitem{GF4} I. Sakalli and O. A. Aslan, Astropar. Phys. \textbf{74}, 73
(2016).

\bibitem{GF5} G. Panotopoulos and A. Rincon, Phys. Rev. D \textbf{96},
025009 (2017).

\bibitem{GF6} C-Y. Zhang, P-C. Li, and B. Chen, Phys. Rev. D \textbf{97},
044013 (2018).

\bibitem{BSs} J. Kunz, \textit{Black Holes In Higher Dimensions (Black
Strings And Black Rings)}, The Thirteenth Marcel Grossmann Meeting: pp.
568-581 (2015); doi.org/10.1142/9789814623995\_0027.

\bibitem{BSs2} R. Gregory and R. Laflamme, Nucl. Phys. B \textbf{428}, 399
(1994).

\bibitem{BSs3} R. Gregory and R. Laflamme, Phys. Rev. Lett. \textbf{70},
2837 (1993).

\bibitem{BS1} I. R. Klebanov and S. D. Mathur, Phys. Nucl. Phys. B \textbf{%
500}, 115 (1997).

\bibitem{BS2} C. G. Callan, S. S. Gubser, I. R. Klebanov, and A. A.
Tseytlin, Nucl. Phys. B \textbf{489}, 65 (1997).

\bibitem{BS3} M. Cvetic and F. Larsen, Phys. Rev. D \textbf{56}, 4994 (1997).

\bibitem{BS4} I. R. Klebanov and M. Krasnitz, Phys. Rev. D \textbf{55},
R3250 (1997).

\bibitem{BS5} S. S. Gubser and I. R. Klebanov, Nucl. Phys. B \textbf{482},
173 (1996).

\bibitem{lem1} J. P. S. Lemos, Class. Quant. Grav. \textbf{12}, 1081 (1995).

\bibitem{lem2} J. P. S. Lemos, Phys. Lett. B \textbf{352}, 46 (1995).

\bibitem{santos} N. O. Santos, Class. Quant. Grav. \textbf{10}, 2401 (1993).

\bibitem{lem3} J. P. S. Lemos and V. T. Zanchin, Phys. Rev. D \textbf{54},
3840 (1996).

\bibitem{lem4} R. G. Cai and Y. Z. Zhang, Phys. Rev. D \textbf{54}, 4891
(1996).

\bibitem{ahmed} J. Ahmed and K. Saifullah, Eur. J. C \textbf{77}, 885 (2017).

\bibitem{dof} M. Cavaglia, Phys. Lett. B \textbf{569}, 7 (2003).

\bibitem{Habib} S. H. Mazharimousavia and M. Halilsoy, Eur. Phys. J. Plus 
\textbf{131}, 138 (2016).

\bibitem{mytachyon} A. Sen, JHEP \textbf{0204}, 048 (2002).

\bibitem{Feinberg}  G. Feinberg, Phys. Rev. \textbf{159}, 1089 (1967). 

\bibitem{Feinberg2} G. Feinberg, Phys. Rev. D \textbf{17}, 1651 (1978).

\bibitem{immass} Y. Aharonov, A. Komar, and L. Susskind, Phys. Rev. \textbf{%
182}, 1400 (1969). 

\bibitem{KK} Th. Kaluza, Sitz. Preuss. Akad. Wiss. Berlin (Math. Phys.) 966
(1921); O. Klein, Z. Phys. \textbf{37}, 895 (1926).

\bibitem{PLB} A. Davidson and D. A. Owen, Phys. Lett. B \textbf{177}, 77
(1986).

\bibitem{Horowitz} G. Horowitz, JHEP \textbf{0508}, 091 (2005).

\bibitem{Born} M. Born and L. Infeld, Proc. Roy. Soc. Lond. A \textbf{144},
425 (1934).

\bibitem{Wu} T.T. Wu, C.N. Yang, i\textit{n Properties of Matter Under
Unusual Conditions, edited by H. Mark and S. Fernbach}, (Interscience, New
York, 1969).

\bibitem{Fernando} S. Fernando, Gen. Rel. Grav. \textbf{37}, 461 (2005).

\bibitem{Flux} C. Ding, S. Chen, and J. Jing, Phys. Rev. D \textbf{82},
024031 (2010).

\bibitem{SurGrav} R. M. Wald, \textit{General Relativity} (The University of
Chicago Press, Chicago and London, 1984).

\bibitem{BY1} J. D. Brown and J. W. York, Phys. Rev. D \textbf{47}, 1407
(1993).

\bibitem{BY2} S. Bose and N. Dadhich, Phys.Rev. D \textbf{60}, 064010 (1999).

\bibitem{BY3} S. Chakraborty and N. Dadhich, JHEP \textbf{12}, 003 (2015).

\bibitem{Steg} M. Abramowitz and I. A. Stegun, \textit{Handbook of
Mathematical Functions} (Dover, New York, 1965).

\bibitem{Lay} S. Y. Slavyanov and W. Lay, \textit{Special Functions: A
Unified Theory Based on Singularities} (Oxford Mathematical Monographs, New
York, 2000).

\bibitem{Kanti} P. Kanti, J. Grain and A. Barrau, Phys. Rev. D \textbf{71}%
,104002 (2005).

\bibitem{tcs} W. G. Unruh, Phys. Rev. D \textbf{14}, 3251 (1976).

\bibitem{chandra} S. Chandrasekhar, \textit{The Mathematical Theory of Black
Holes} (Oxford University Press, New York, 1983).
\end{thebibliography}
\end{document}